\documentclass{emulateapj}
\newcommand{\alfven}{Alfv\'{e}n}
\newcommand{\alfvenic}{Alfv\'{e}nic}
\newcommand{\pref}{\protect\ref}

\slugcomment{In Press Ap. J. [March 2009]}
\begin{document}

\shorttitle{Coronal Seismology}
\shortauthors{Tomczyk \& McIntosh}
\title{Time-Distance Seismology of the Solar Corona with CoMP}
\author{Steven Tomczyk, Scott W. McIntosh}
\affil{High Altitude Observatory,\\ National Center for Atmospheric Research,\\ P.O. Box 3000, Boulder, CO 80307}
\email{tomczyk@ucar.edu, mscott@ucar.edu}

\begin{abstract}
We employ a sequence of Doppler images obtained with the Coronal Multi-channel Polarimeter (CoMP) instrument to perform time-distance seismology of the solar corona. We construct the first $k-\omega$ diagrams of the region. These allow us to separate outward and inward propagating waves and estimate the spatial variation of the plane-of-sky projected phase speed, and the relative amount of outward and inward directed wave power. The disparity between outward and inward wave power and the slope of the observed power law spectrum indicate that low-frequency \alfvenic{} motions suffer significant attenuation as they propagate, consistent with isotropic MHD turbulence.
\end{abstract}


\section{Introduction}
The Sun's corona is the primary source of the disturbances to the Earth's environment known as space weather. Despite its relevance, our knowledge about fundamental physical parameters in the solar corona is still quite rudimentary. The measurement of coronal temperatures, densities and abundances is difficult. And although magnetism dominates the force balance in the corona, there exist very few direct measurements of the strength and orientation of coronal magnetic fields \citep[see, e.g.,][]{Brosius2006, Lin2004, Tomczyk2008}. Consequently, the mechanisms responsible for heating the corona, driving the solar wind, and initiating coronal mass ejections remain poorly understood. 

Over the past decade a great deal of progress has been made in observing and interpreting the signatures of magneto-hydrodynamic (MHD) waves in the corona. Various types of MHD waves have been identified including fast mode shocks seen as EIT waves \citep[e.g.,][]{Thompson}, slow magneto-acoustic (MA) waves in polar plumes \citep[][]{DeForest} and coronal loops \citep[e.g.,][]{DeMoortel, Marsh} and fast MA kink waves \citep[e.g.,][]{Asch1999, Nak1999}. The last of these has been the focus of much recent work where {\em Transition Region and Coronal  Explorer} \citep[TRACE;][]{Handy1999} measurements of impulsively excited transverse coronal loop displacements have been used to constrain the strength of the coronal magnetic field and other atmospheric parameters \citep[][]{Asch1999, Nak1999, Asch2002, Sch2002}. Such investigations of the coronal magnetic structure and plasma environment, in the limit of MHD, has been dubbed coronal seismology \citep[e.g.,][]{Asch2003, Nak2005, Ban2007}. Observable loop displacement oscillations are relatively rare and only a few dozen loops have been analyzed seismologically. 

Recently, we discovered spatially and temporally ubiquitous waves in the solar corona in the time series of Doppler images taken with the Coronal Multi-channel Polarimeter (CoMP) instrument \citep{Tomczyk2007}. These data present the exciting possibility to greatly widen the application of coronal seismology. Analysis revealed propagating waves with an amplitude of $\sim$300 m/s rms (1 km/s peak-to-peak) and peak power at a period of 5 minutes with \alfvenic{} phase speeds ($\sim$1 Mm/s) and with very small associated intensity fluctuations. These waves were interpreted by \citet{Tomczyk2007} as propagating \alfven{} waves. 

These ubiquitous propagating waves eluded earlier investigators for several reasons. First, these waves do not have appreciable intensity fluctuations so they have not been seen with intensity imaging instruments. Also, a velocity amplitude of 1~km/s at a period of 5~minutes corresponds to a loop displacement amplitude of 48~km which is nearly an order of magnitude smaller than the 365~km (0.5\arcsec) pixel size of TRACE and nearly two orders of magnitude smaller than the $\sim$2000 km pixel size of SOHO/EIT \citep[2.6\arcsec;][]{Boudine1995}. The observation of these oscillations as loop displacements is probably beyond the capability of current instruments. In addition, Doppler imaging measurements with the required sensitivity level, spatial extent and temporal cadence are difficult (or impossible) to obtain with existing slit scanning spectrograph instruments. For example, the SOHO/SUMER instrument \citep[][]{Wilhelm1995} has a spectral sampling of approximately 10~km/s (43~m\AA) per pixel in first order; observation of these ubiquitous coronal waves with SUMER would require line centroiding to considerably better than 0.1~pixel. Recent analysis of spectra from the EIS instrument \citep[][]{Culhane2007} on {\em Hinode} \citep[][]{Kosugi2007} show a portion of the corona oscillating at a period very near 5 minutes with a peak-to-peak amplitude of 2~km/s \citep[][]{Van2008b}, indicating that these ubiquitous waves may be marginally observable with EIS. Ground-based spectroscopic observations with a coronagraph showed clear evidence of oscillations, but their identification as propagating \alfvenic{} disturbances was inconclusive \citep[][]{Sakurai}. 

In this paper we extend the analysis of the same dataset used in the work by \citet{Tomczyk2007}. They noted the clear presence of outward and inward propagating \alfven{} waves in the corona, but no attempt was made to separate them. We present here a method to separate outward and inward propagating waves through the construction and analysis of coronal time series projected onto the paths of wave propagation. These time-distance diagrams are suitable for spatio-temporal Fourier analysis allowing us to build the first $k-\omega$ diagnostic diagrams of the corona. As we will show, these allow significant improvement in the determination of the phase speed of the waves, which enhances our ability to perform coronal seismology. 

In the following section we will briefly discuss the CoMP instrument and the observations analyzed in this paper. In Sect. 3 we discuss the data analysis techniques that we employ to derive the time-distance diagnostic measures. In Sect. 4 we discuss the implications of our results and place them in context with previous work. We conclude with a discussion of the potential for future observations to advance the field of coronal seismology.


\begin{figure}
\epsscale{1.1}
\plotone{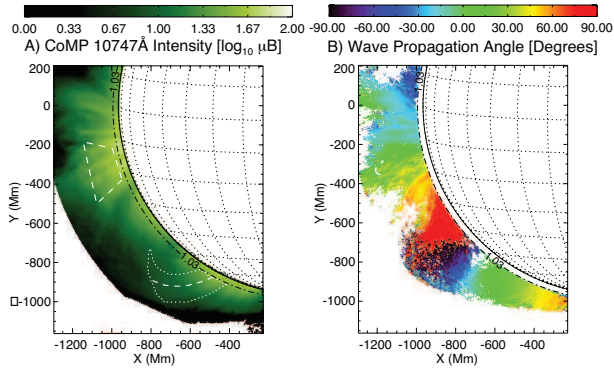}
\caption{The \ion{Fe}{13} 10747 \AA{} intensity (panel A) and derived wave propagation angles (panel B) in the region of the corona studied by CoMP on 2005 October 30. On panel A we show, for later reference, the active loop region (dashed trapezoid) over which we have summed Fourier power spectra of the Doppler velocity to construct Fig.~\pref{f2}, the sample path used to determine the wave diagnostics (dashed curved line;) and the surrounding (dotted) region to examine the spatially averaged $k-\omega$ diagram studied in Fig.~\pref{f4}. The wave propagation angles are measured counter-clockwise from due East. \label{f1}}
\end{figure}

\section{Instrument and Observations}\label{obs}
The CoMP instrument is a combination polarimeter and narrowband tunable filter that can measure the complete polarization state in the vicinity of the \ion{Fe}{13} infra-red coronal emission lines at 10747 \AA{} and 10798 \AA{} and the chromospheric 10830 \AA{} \ion{He}{1} line with a 2.8 $R_\odot$ field-of-view and a sampling of 4.5~arcseconds/pixel. The bandpass of the filter is tuned electro-optically with a bandwidth of~1.3 \AA{} (FWHM). CoMP was deployed behind the 20-cm aperture Coronal One Shot coronagraph \citep{Smartt1981} on the equatorial spar of the Hilltop facility at the Sacramento Peak Observatory of the National Solar Observatory in Sunspot New Mexico. 

As in \citet{Tomczyk2007}, the observations used in this paper were obtained on Oct 30, 2005 and were restricted to the  linear polarization (Stokes I, Q and U only) of the 10747 \AA{} \ion{Fe}{13} line, sampling three wavelengths (10745.2, 10746.5, and 10747.8 \AA) across the line. The exposure time for the images was 250~ms and image groups of full polarization and wavelength information were obtained at a cadence of approximately 28.7 seconds. The analysis presented in this paper is restricted to the first 350 image sets due to their superior quality, covering 2.8 hours starting at 14.261~UT. We selected a sub-array on the east limb for further analysis (see, Panel A of Fig.~\pref{f1}) which contains both active region loops and a coronal cavity.

For each pixel in the selected sub-array we fit a Gaussian profile to the intensity vs. wavelength and obtained estimates for the central intensity, central wavelength, and linewidth. The line-of-sight (LOS) component of the velocity was derived from the Doppler shift of the central wavelength, and we computed the plane-of-sky direction of the magnetic field ($\phi$) using the measured Stokes Q and U at line center and the relationship
\begin{equation}\label{eq1}
\phi = \frac{1}{2} tan^{-1}(\frac{U}{Q}).
\end{equation}

\begin{figure}
\epsscale{1.1}
\plotone{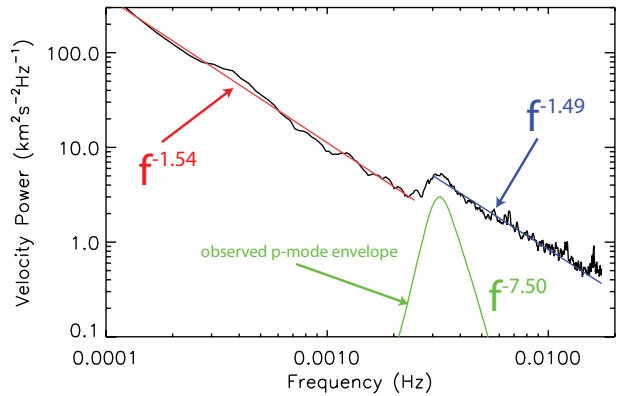}
\caption{The log-log Fourier spectrum of the measured Doppler velocity averaged over the region shown in panel A of Fig.~\pref{f1}. Also shown are the low- (red), high-frequency (blue) power-law fits and the photospheric p-mode spectrum (green). \label{f2}}
\end{figure}

In Fig.~\pref{f2} we show a power spectrum of the coronal LOS velocity fluctuations averaged over the area above the active region to the South of the equator in Fig.~\pref{f1}. It shows that the power spectrum has a peak at a frequency of 3.2 mHz; above 3.2 mHz the signal is dominated by oscillatory power. The spectrum of photospheric p-modes computed by averaging all intermediate degree p-mode spectra with spherical harmonic degrees between 0 and 199 observed on one day with the LOWL instrument \citep{Tomczyk1995} is also shown on the plot\footnote{The p-mode spectrum shown has been filtered to remove power below 1~mHz and been scaled to fit on the plot.}. We notice that both the low- and high-frequency portions of the spectrum obey power-law scaling. Fitting linear relationships to the log-log coronal wave power spectrum in the low- (0.1 $\ge$ f $\ge$ 2.5~mHz) and high-frequency (3 $\ge$ f $\ge$ 10~mHz) ranges we obtain values of the slopes of -1.54 $\pm$ 0.02 and -1.49 $\pm$ 0.02, respectively, where we quote the 1-$\sigma$ errors in the fit gradients. The relevance of the power-law scaling in the wave spectrum will be discussed below (see, Sect.~\pref{results}). 

\section{Analysis 1. Determination of Wave Propagation Direction}\label{anal1}

The first step of the time-distance analysis is to determine the direction of wave propagation. The analysis method used here is described in detail in \citet{McIntosh2008} and is a modification of the method employed in \citet{Tomczyk2007}. We determine the coherence (the frequency space equivalent of the cross-correlation) of the velocity time series of each pixel with its surrounding pixels. These coherence maps reveal an ``island'' of high coherence (defined as coherence greater than 0.5). We define the direction of wave propagation to be aligned with the island and compute the orientation of the island to be along a line which minimizes the sum of perpendicular distance from the points in that island to the line. The method employed in \citet{Tomczyk2007} differed in that it determined the time domain cross-correlation and fit the orientation of the islands using a simple linear least-squares fit. We subsequently determined that the original method failed to determine the wave angle well when the wave angle was highly inclined. 

The map of wave angles is shown in Panel B of Fig.~\pref{f1} along with the coronal \ion{Fe}{13} intensity. Comparing this angle map with the corresponding  panel of \citet{Tomczyk2007} (Fig.~4 panel B; showing angles with errors of less than $5^\circ$) we see that while the general spatial pattern is the same, the new approach has been able to reach considerably higher inclinations of the wave trajectory. 

\begin{figure}
\epsscale{1.0}
\plotone{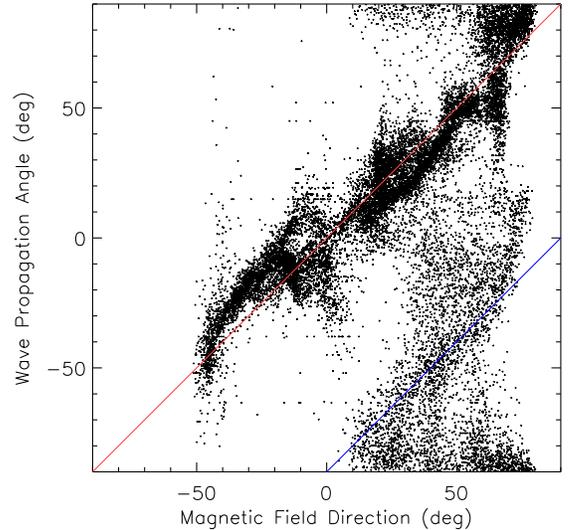}
\caption{Scatter plot of the measured magnetic field direction \citep[panel F of Fig.~1 in][]{Tomczyk2007} and the derived wave propagation angles (panel B of Fig.~\pref{f1}). The solid red line indicates one-to-one correspondence while the blue line designates the material subject to the Van Vleck ambiguity that is offset by $90^\circ$. \label{f3}}
\end{figure}

In Figure~\pref{f3} we show the comparison of the wave direction with the directly measured magnetic field direction from the linear polarization measurements and Eq.~\pref{eq1}. Due to the above algorithm refinement, the correlation between the wave direction and magnetic field orientation is improved over that presented in \citet{Tomczyk2007}. The magnetic field direction obtained from linear polarization measurements is subject to the well known $90^\circ$ Van Vleck ambiguity \citep[][]{vanvleck, House}. This ambiguity occurs when the angle between the incident radiation field and the magnetic field direction passes through an angle of $54.6^\circ$. The ambiguity is evident in the figure as the locus of points that lie parallel to but $90^\circ$ below the correlation line. We define the points which are subject to the Van Vleck ambiguity as those that are more than $45^\circ$ away from the correlation line and find that about 20\% of the points in the figure are subject to the Van Vleck ambiguity. Neglecting these points, the correlation coefficient between the wave propagation direction and the magnetic field direction is 0.91. This confirms our previous assertion \citep[][]{Tomczyk2007} that these waves are propagating along magnetic field lines imbedded in the corona. We anticipate that the Van Vleck effect will, in the future, provide a strong constraint on the magnetic field geometry and the location of the emitting region along the line-of-sight.

\section{Analysis 2. Construction of Space-Time and $k-\omega$ Diagrams}\label{anal2}

\begin{figure}
\epsscale{1.1}
\plotone{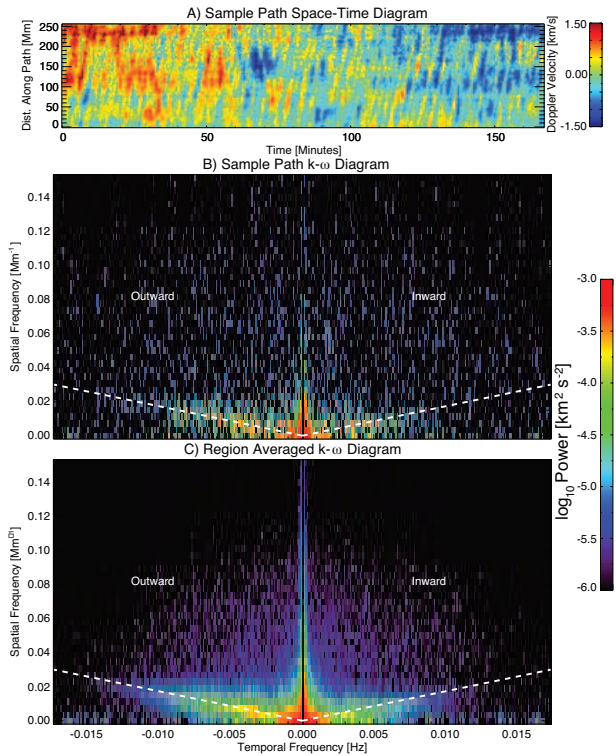}
\caption{CoMP space-time diagram for the sample wave path traced out in Fig.~\pref{f1} (panel A), the $k-\omega$ diagram for that space-time diagram (panel B) and the average $k-\omega$ diagram for the region around the sample wave path (panel C). \label{f4}}
\end{figure}

We employ the map of wave directions computed above to trace the wave paths in the corona, starting from any location in the field of view. This is accomplished by stepping a small amount (typically one pixel) from a starting point in the direction of wave propagation and then interpolating a new direction at that point and repeating the process. An example wave path is shown in Fig.~\pref{f1} with a length of 80 pixels ($\sim$250 Mm). Next, we compute the velocity for each point along the wave path for each velocity image in the time series and stack them into a space-time diagram. We use a cubic interpolation algorithm to map the velocities in the original data cube onto the wave path. An example space-time diagram is shown in panel A of Fig.~\pref{f4} for the wave-path structure traced in Fig.~\pref{f1}. We see outward and inward propagating waves as linear structures in the time series that are inclined to the right and left, respectively. The angle of the structure to the vertical in the space-time diagram gives a measure of the wave's phase speed. 

The $k-\omega$ diagram of the path is the Fourier power spectrum of the space-time diagram, see Panel B of Fig.~\pref{f4}. It shows ridges of power in both positive and negative $\omega$ that correspond to outward and inward propagation of the waves, respectively. We can also see that the wave power in each direction is not equal. In this case the outward directed wave power is about a factor of 2 greater than the inward power. The ridges of power in positive and negative $\omega$ appear to have the same inclination, i.e., the same phase speed. The dashed lines shown in the panel correspond to phase speeds of 600~km/s which is significantly lower than the 1.2~Mm/s reported by \citep{Tomczyk2007} for this region. Also visible in the panel is a vertical ridge of power at very low values of $\omega$. This is due to the large amount of low frequency velocity power consistent with the $f^{-3/2}$ spectrum shown in Fig 2. The example single-path $k-\omega$ diagram can be enhanced by averaging the diagrams of neighboring wave-paths. In Panel C of Fig.~\pref{f4} we have averaged the $k-\omega$ diagrams for the paths in the dashed region surrounding the single path shown in Fig.~\pref{f1}. We see that this region-averaged $k-\omega$ diagram shows the same properties as that of Panel B of Fig. \pref{f4}.
 
\begin{figure}
\epsscale{1.1}
\plotone{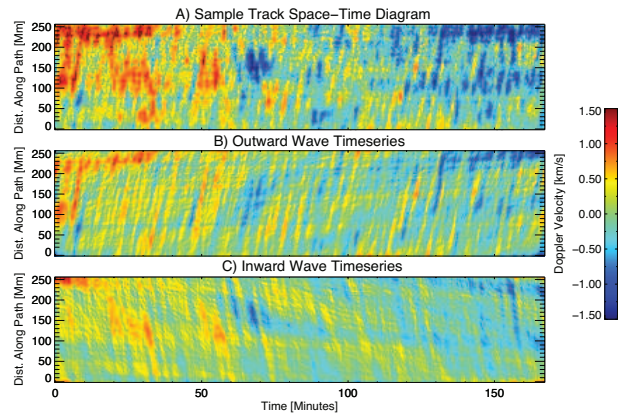}
\caption{Sample CoMP space-time diagram for the sample structure traced out in Fig.~\pref{f1} (panel A), the corresponding $k-\omega$ filtered outward (negative-$\omega$; panel B) and inward (positive-$\omega$; panel C) space-time diagrams. \label{f5}}
\end{figure}

Fitting the ridges in these $k-\omega$ diagrams to extract the phase speed of the waves in each direction is not straightforward due to the large amount of low spatial and temporal frequency power present, especially for the single-path timeseries, and so we must find another method. Taking the simplest possible approach, we compute the space-time diagram of outward and inward waves by masking the positive and negative frequency halves of the complex $k-\omega$ transform separately and compute the inverse transform. The outward and inward filtered space-time diagrams corresponding to the example of Fig. \pref{f4} are shown in panels B and C of Fig.~\pref{f5}. The outward and inward timeseries show a few key properties. First, the correspondence between the wave patterns in the filtered (panels B and C) and the unfiltered (original, panel A) timeseries is clear with the vertical stripes oriented as we would expect, with the outward waves inclined to the right and the inward waves inclined to the left. The phase speed of the waves, indicated by the gradient of the stripes, appears to be approximately constant over the duration of the time series in each case.

\begin{figure}
\epsscale{1.1}
\plotone{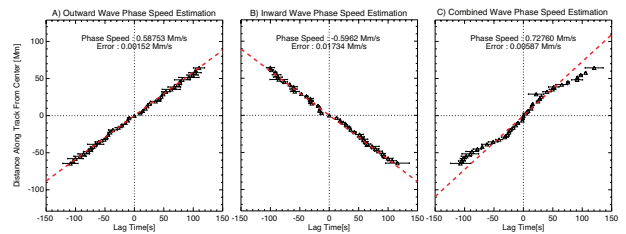}
\caption{Cross-correlation phase-speed determination for the reconstructed (outward - panel A; inward - panel B) and the unfiltered (panel C) space-time diagrams. The phase-speed in each case is the gradient of the linear-fit to the wave lag at different positions along the wave path. \label{f6}}
\end{figure}

Using the filtered timeseries we are able to measure the wave phase speed along the path by cross-correlating the timeseries at the center of the path with those across the entire length of the path. We determine the lag of the cross-correlation functions [where cc $>$ 0.5] with a parabola (the peak of the parabola determines the lag of the wave) and obtain the phase speed by fitting the slope of the observed lag with position along the path. Examples of the cross-correlation functions and resulting space-lag plots for the outward, inward and original unfiltered timeseries are shown in Fig.~\pref{f6}. We see that the phase speeds of the waves in the outward, inward, and unfiltered timeseries are 587 ($\pm$ 2), 596 ($\pm$ 20) and 728 ($\pm 6$) km/s, respectively. We note that the outward and inward phase speeds are consistent, but the phase speed of the composite is much higher. This discrepancy is understandable in terms of linear wave theory by considering a situation where waves are traveling in opposite directions on the same structure with the same speed but with different amplitudes. When the amplitudes are equal the combination of the two traveling waves is a standing wave; as the relative amplitude increases in one direction or the other the net wave will appear to travel in the direction of dominant amplitude. Since the work of \citet{Tomczyk2007} did not separate outward and inward waves but instead analyzed the composite time series, that work severely overestimated (by a factor of 2) the phase speeds except where the wave power was dominated by either the outward or inward propagating component - the phase speeds shown here supersede those in \citet{Tomczyk2007}.

\begin{figure*}
\epsscale{1.0}
\plotone{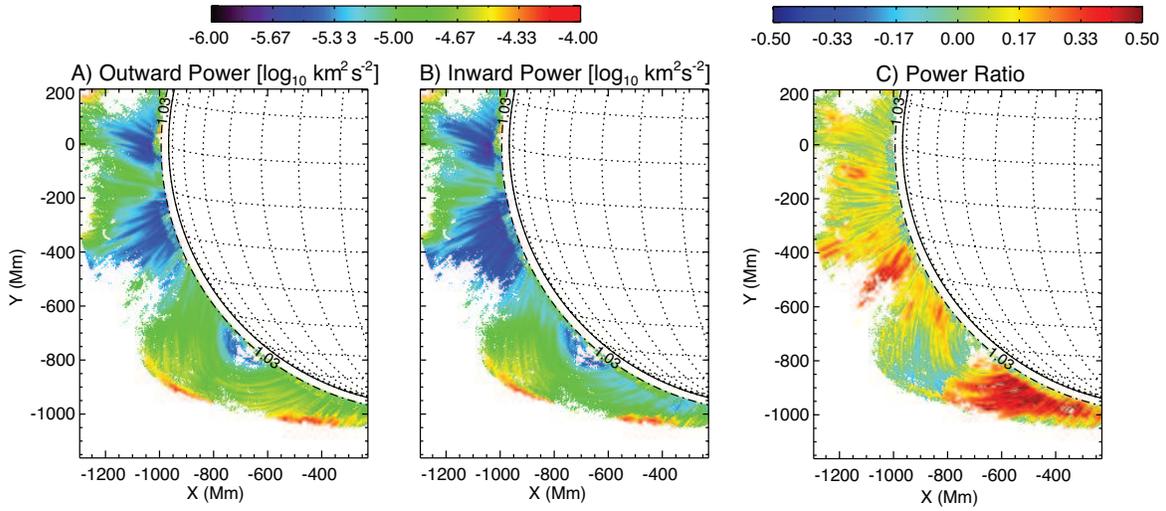}
\caption{Maps of the outward ($P_{out}$; panel A), inward ($P_{in}$; panel B) directed wave velocity power and their ratio ($R = (P_{out}-P_{in})/(P_{out}+P_{in})$; panel C).\label{f7}}
\end{figure*}

\begin{figure*}
\epsscale{1.0}
\plotone{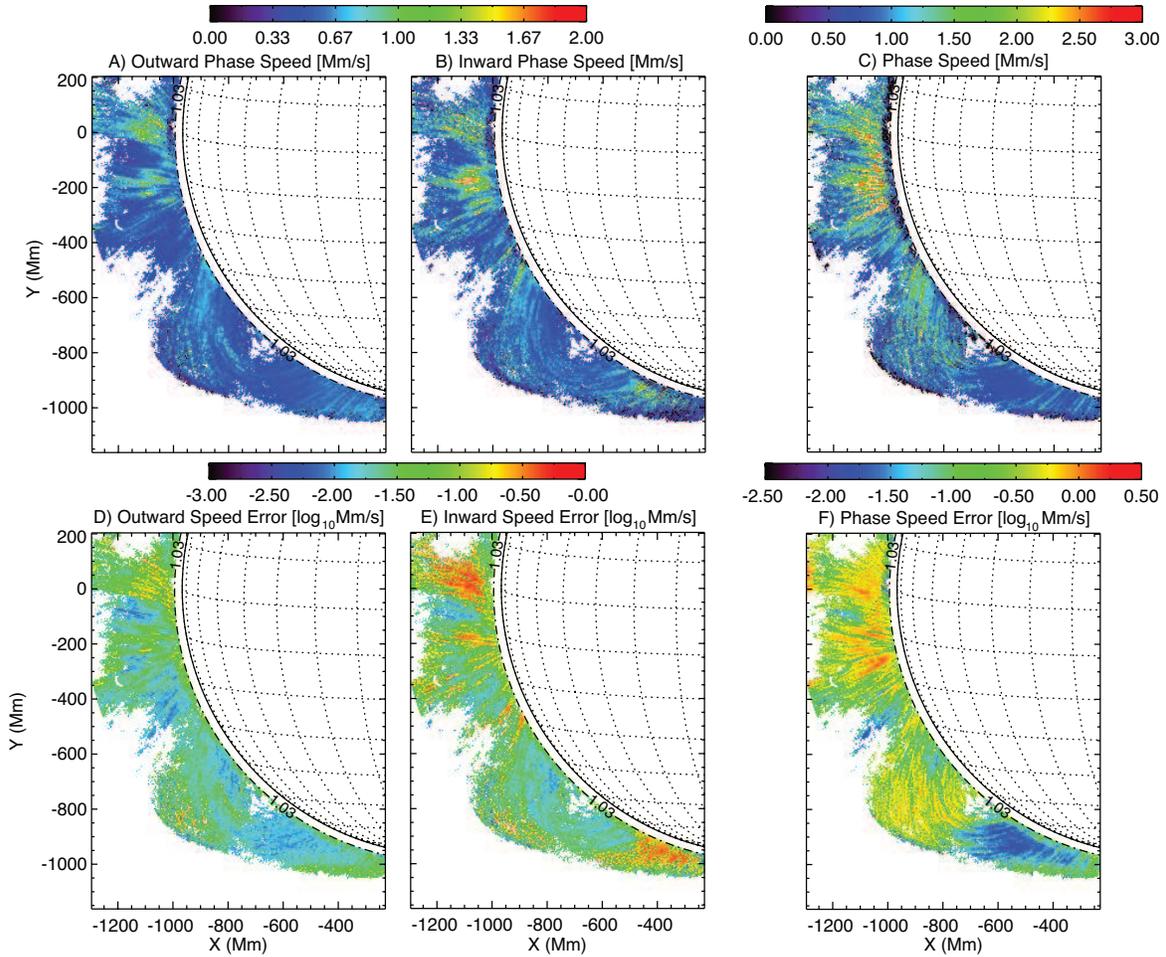}
\caption{Maps of the outward (panel A), inward (panel B) directed and unfiltered (panel C) wave phase speeds and respective errors (panels D, E and F).\label{f8}}
\end{figure*}

As in \citet{Tomczyk2007} we can repeat all the steps of the analysis pixel by pixel over the field-of-view shown in Fig.~\pref{f1}. In doing this, we have computed wave paths of 31 pixels in length ($\sim$100 Mm) centered on the pixel of interest. We chose to use a short path for this step of the analysis to keep the wave paths from extending appreciably beyond the field-of-view. The value of 31 pixels was the shortest length that gave reliable results. From these wave paths, we compute filtered space-time diagrams and extracted the phase speed in each, as well as that in the composite timeseries. We also compute the power in the outward and inward portions of the $k-\omega$ diagram for each path. Results of this analysis are presented in Figs.~\pref{f7} and~\pref{f8}. The panels of Fig.~\pref{f7} show the resulting maps of outward and inward wave power as well as their ratio [$\frac{P_{out} -  P_{in}}{P_{out} + P_{in}}$] from left to right while the panels of Fig.~\pref{f8} show the respective phase speed maps in the three cases (outward - panel A; inward - panel B; unfiltered space-time diagram - panel C) and their respective errors. The power maps show excess power at the highest parts of the field-of-view. This is due to the decreasing coronal intensity at these heights and correspondingly higher noise levels there. The power ratio map shows that the corona is dominated by outward propagating waves except in a small region on the east side of the coronal cavity where inward wave propagation slightly dominates. Due to the dominance of outward wave power, the inward phase speed is generally less well determined than the outward phase speed and that is indicated by the larger errors seen.

\section{Discussion}\label{results}

The observed peak of the spectrum of coronal velocity perturbations in Fig.~\pref{f1} shows an exact coincidence with the peak of the p-mode power spectrum. From this we conclude that the ubiquitous coronal waves observed in this frequency range are driven by solar p-modes. Recent work supports the contention that p-modes can be efficiently transported into the corona via coupling with magnetic fields rooted in the photosphere \citep[][]{depont, deros, Hindman}. 

We observe a power-law spectrum in the Doppler velocity over our entire frequency range, with excess power in the high-frequency (f $>$ 2 mHz) oscillatory region of the spectrum. A power-law spectrum has been widely observed in fluctuations of the solar wind with a slope between -5/3 and -3/2 which has been interpreted as evidence of a turbulent cascade \citep[see, e.g.,][]{Goldstein}. A slope of -3/2 is predicted for isotropic incompressible MHD turbulence \citep{Kraichnan}. The CoMP observations reveal a slope of -3/2 in both the low and high frequency regimes consistent with a turbulent cascade.

It is noteworthy that the -3/2 slope of the high frequency (f $>$ 3 mHz) tail of the coronal velocity spectrum is in contrast to the photospheric velocity spectrum in this frequency range which declines as $\sim f^{-7.5}$. This is evidence that the high frequency coronal spectrum is not merely imprinted by the photospheric spectrum. Either some mechanism modifies the photospheric spectrum as waves propagate into the corona or the coronal spectrum is modified {\it in situ}.

We observe wave phase speeds that are approximately constant over the duration of the observations. This implies that the magnetic structures supporting the waves and the driving mechanism are likewise stable over this time scale. We note the large differences between inward and outward wave power, even along closed loop structures like the coronal cavity region. This can only be possible if the waves are damped appreciably {\it in situ}. The approximate time scale for damping must be of the order of the travel time along the loop, which amounts to just a few wave periods. This is consistent with the observed damping times of coronal loop oscillations \citep{Asch2002}. When considering the efficiency of these waves in heating the corona, or driving the solar wind \citep[][]{DePontieu2007}, we must explore this damping process further, consider the origins of the observed large non-thermal line widths in the corona and exploit the developing self-consistent theories which involve the effective dissipation of low-frequency \alfvenic{} motions \citep[e.g.,][]{Cranmer2005, Suzuki2005, Cranmer2007, Verdini2007}. This will be the subject of a forthcoming paper. 

As we have noted above, \citet{Tomczyk2007} interpreted the waves seen in the CoMP data as \alfven{} waves. This conclusion is consistent with the observations that the measured phase speeds of the waves are much larger than the sound speed, that the waves propagate along field lines and that the waves exhibit very small intensity perturbations. \citet{Van2008a} and \citet{Hindman} suggest that these waves are more likely fast MA kink modes. The classical \alfven{} wave formulation \citep{alf} assumes a uniform magnetic field in a uniform density environment. MA kink modes arise in a mathematical formulation that assumes a cylindrical loop geometry with a density jump between the inside and outside of the loop \citep{Edwin}. In such a formulation, the \alfven{} mode is torsional and will not be detected in observations where the loops are not spatially resolved, which is the case for the CoMP observations. In fact, coronal loops have yet to be conclusively resolved by {\em any} instrument. We admit the possibility that the waves seen in \citet{Tomczyk2007} and this paper may, in fact, be fast MA kink modes. However, we have no information about the density structure of the coronal features studied here, so an unambiguous interpretation in terms of a specific geometry is unwarranted at this time. Characterization of these waves as the more generic \alfven{} waves is in our opinion somewhat less presumptuous.

Further, in the MA kink mode interpretation, under the usual assumption of a uniform magnetic field inside and outside the loop, the phase speed, $V_{phase}$, is approximately given by the kink speed, $V_{kink}$, and is related to the \alfven{} speed inside the loop, $V_{Ai}$ ($= {B_i \over \sqrt{\mu_i \rho_i}}$) by the relation
\begin{equation}
V_{phase} \sim V_{kink} = V_{Ai} \sqrt{2 \over {1+ {\rho_o / \rho_i}}}~,
\end{equation}
where $\rho_i$ and $\rho_o$ are the densities inside and outside the loop, respectively and $\mu_i$ is the mean molecular weight inside the loop \citep[see, e.g.,][]{Nak2005}. The maximum difference between the kink speed and \alfven{} speed in going from a uniform density environment to a loop in an evacuated environment is a factor of $\sqrt{2}$. This corresponds to the maximum systematic error in the inferred value of the magnetic field due to an interpretation in terms of \alfven{} waves or MA kink mode waves. Recent spectroscopic observations with EIS were used to detect a coronal oscillation and to map the electron density variation across a coronal loop \citep{Van2008b}. That study indicated a 20\% density enhancement of the loop over its surroundings which gives rise to only a 5\% difference in the value of the inferred magnetic field using either an \alfven{} wave or MA kink mode interpretation. The investigation of \citet{Young2008} into \ion{Fe}{12} and \ion{Fe}{13} electron density diagnostics with EIS indicates discrepancies of up to 0.5 in the log of the inferred electron density that are due, principally, to uncertainties in the atomic models employed by the CHIANTI database \citep[][]{Dere1997}. While we acknowledge that it is important to determine the actual type of wave responsible for these ubiquitous coronal oscillations and to infer physically meaningful information from them, we suggest that the uncertainty in the magnetic field inferred from observed wave phase speeds is dominated by uncertainty in the density to a degree that the difference between the kink and \alfven{} speed is small in comparison. In the future, CoMP will complement all wave sequences with measurements of circular polarization (Stokes V) to directly measure the LOS magnetic field strengths \citep[see, e.g., Fig.~8 of][]{Tomczyk2008}, and density diagnostics of the \ion{Fe}{13} line pair \citep[see, e.g.,][]{penn,singh} in an effort to address this issue.


Finally, as in \citet{Tomczyk2007}, we can estimate the energy carried by the ubiquitous coronal waves as: $F_{W} = \rho \langle v^2 \rangle V_{phase}$, where $\rho$ is the density, $v$ is the velocity amplitude, and $V_{phase}$ is the phase speed of the waves. Assuming a typical range of electron densities of $10^{8}-10^{9}$~cm$^{-3}$, we find $\rho \sim 2 \cdot 10^{-16} - 2 \cdot 10^{-15}$g~cm$^{-3}$; also, $v \sim 300$~m/s and the median phase speed over the field is 549.2~$\pm$28.6~km/s. Then, the flux of energy propagating in the observed waves is $F_{W}$ = 10-100 erg~cm$^{-2}$s$^{-1}$ compared with the estimate of $3 \cdot 10^{5}$erg~cm$^{-2}$s$^{-1}$ required to balance the radiative losses of the quiet solar corona \citep{with}. We concur with our earlier estimate \citep{Tomczyk2007} that the energy carried in the {\em resolved} ubiquitous coronal waves is insufficient to heat the solar corona by at least three to four orders of magnitude. Calculation of the energy carried by these waves assuming they are MA kink modes could reduce the energy estimate by as much as an order of magnitude \citep{Van2008a} but leaves our basic conclusion intact. As in our earlier paper, we caution that this calculation includes only the waves that we can spatially resolve with our instrument and is therefore an underestimate. The ability of these waves to heat the corona remains an open question.

\section{Conclusion}

Precision Doppler imaging of the corona with the CoMP instrument has revealed oscillations which are ubiquitous in space and time. In this paper, we have exploited the ubiquity of the \alfvenic{} motions to construct the first space-time and $k-\omega$ diagrams of the corona. These allow us to separate outward and inward propagating waves and obtain a precise estimate of the spatial variation of the plane-of-sky projected phase speed. The separation of inward and outward waves is critical to the precise estimation of the phase speed and represents a significant improvement over the analysis of \citet{Tomczyk2007}. The significant disparity between inward and outward wave power over most of the corona leads us to conclude that the dissipation time scale for these waves is roughly equal to the time of wave propagation across the loops. This is consistent with a picture of the corona where low-frequency \alfvenic{} motions suffer from significant damping as they propagate along magnetic structures. The power law spectra with a slope of -3/2 that we observe is consistent with MHD turbulence. However, other dissipation processes such as wave mode conversion \citep[e.g.][]{Melrose} cannot be ruled out. 

We have shown that observations of this type can provide precise phase speed information which is an essential ingredient to the broad practical application of coronal seismology. However, since we lack information on the density of the corona in the region we have studied, we can not perform real coronal seismology with this data set at this time. We suggest that the lack of precise density information in coronal structures is the largest obstacle to the widespread application of coronal MHD seismology to the study of coronal magnetism. Current uncertainties in the density distribution in the corona dominate even over uncertainty due to the assumption about the nature of the wave being observed. 

Coronal seismology measurements constrain the plane-of-sky projected phase speed and therefore the transverse component of the magnetic field which is complementary to the coronal Zeeman Effect measurements \citep[e.g.][]{Lin2004, Tomczyk2008} which constrain the LOS component of the magnetic field. The combination of seismology and Zeeman diagnostics then has the potential to constrain the complete {\em vector} magnetic field in the corona. We are currently in the process of relocating the CoMP instrument from Sunspot, NM to Haleakala, HI where we expect the number and quality of coronal observations with the CoMP instrument to be greatly increased.

\acknowledgements 
The authors would like to thank Phil Judge and an anonymous referee for thoughtful comments on the manuscript. The work presented in this paper was supported by the National Science Foundation through the NCAR Strategic Initiative Fund and through HAO base funds, and by the National Aeronautics and Space Administration through grant NNX08AU30G issued by the Living with a Star Targeted Research and Technology Program to SWM. Additional support to SWM came from grants ATM-0541567, NNG06GC89G and NNX08AL22G. The National Center for Atmospheric Research is sponsored by the National Science Foundation.


\begin{thebibliography}{72}
\bibitem[{{\alfven{}} (1942)}]{alf}
\alfven{}, H., 1942, Nature, 150, 405

\bibitem[{{Aschwanden}(2003)}]{Asch2003}
Aschwanden, M.~J. 2003, In: Turbulence, Waves and Instabilities in the Solar Plasma, 
NATO Sci. Series, 124, 215

\bibitem[{{Aschwanden} {et~al.}(1999)}]{Asch1999}
Aschwanden, M.~J., Fletcher, L., Schrijver, C.~J. \& Alexander, D. 1999, \apj, 520, 880

\bibitem[{{Aschwanden} {et~al.}(2002)}]{Asch2002}
Aschwanden, M.~J., de~Pontieu, B., Schrijver, C.~J. \& Title, A.~M. 2002, \solphys, 206, 99

\bibitem[{{Banerjee} {et~al.}(2007)}]{Ban2007}
Banerjee, D., Erde\'lyi, R., Oliver, R., O'Shea, E. 2007, \solphys, 246, 3

\bibitem[{{Bevington} \& {Robinson}(2003)}]{Bevington2003}
Bevington, P.~R. \& Robinson, D.~K. 2003, Data reduction and error analysis for the physical sciences, Boston, MA: McGraw-Hill

\bibitem[{{Delaboudini\`{e}re} {et~al.}(1995)}]{Boudine1995}
Delaboudini\`{e}re, J.-P., et~al. 1995, \solphys, 162, 291

\bibitem[{{Dere} {et~al.}(1997)}]{Dere1997}
Dere, K.~P., et~al. 1997, \aap{}~Supp., 125, 149

\bibitem[{{Brosius} \& {White}(2006)}]{Brosius2006}
Brosius, J.~W., \& White, S.~M. 2006, \apj, 641, L69

\bibitem[{{Cranmer} \& {van Ballegooijen}(2005)}]{Cranmer2005}
Cranmer, S.~R. \& A.~A. van Ballegooijen 2005, \apjs, 156, 265

\bibitem[{{Cranmer}, {van Ballegooijen} \& {Edgar}(2007)}]{Cranmer2007}
Cranmer, S.~R., A.~A. van Ballegooijen \& Edgar, R.~J. 2007, \apjs, 171, 520

\bibitem[{{Culhane} {et~al.}(2007)}]{Culhane2007}
Culhane, J.~L., et~al. 2007, \solphys, 243, 19

\bibitem[{{DeForest} \& {Gurman}(1998)}]{DeForest}
DeForest, C.~E. \& Gurman, J.~B. 1998, \apj, 501, L217

\bibitem[{{De~Moortel}, {Ireland} \& {Walsh}(2000)}]{DeMoortel}
De~Moortel, I., Ireland, J. \& Walsh, R.~W. 2000, A \& A, 355, L23

\bibitem[{{De~Moortel} \& {Rosner}(2007)}]{deros}
De~Moortel, I. \& Rosner, R. 2007, \solphys, 246, 53

\bibitem[{{De~Pontieu}, {Erd\'elyi} \& {De~Moortel}(2005)}]{depont}
De~Pontieu, B., Erd\'elyi, R., \& De~Moortel, I. 2005, \apj, 624, L61

\bibitem[{{De Pontieu} {et~al.}(2007)}]{DePontieu2007}
{De Pontieu}, B., {et~al.} 2007c, Science, 318, 1574

\bibitem[{{Edwin} \& {Roberts}(1983)}]{Edwin}
Edwin, P.M. \& Roberts, B. 1983, \solphys, 88, 179

\bibitem[{{Goldstein} \& {Roberts}(1999)}]{Goldstein}
Goldstein, M.~L., \& Roberts, D.~A. 1999, Physics of Plasmas, 6, 4154

\bibitem[{{Handy} {et~al.}(1999)}]{Handy1999}
{Handy}, B.~N., et~al. 1999, \solphys, 187, 229

\bibitem[{{Hindman} \& {Jain}(2008)}]{Hindman}
Hindman, B.~W. \& Jain, R. 2008, \apj, 677, 769

\bibitem[{{House}(1972)}]{House}
House, L.~L. 1972, \solphys, 23, 103

\bibitem[{{Kraichnan}(1965)}]{Kraichnan}
Kraichnan, R.~H. 1965, Physics of Fluids, 8, 1385

\bibitem[{{Kosugi} {et~al.}(2007)}]{Kosugi2007}
Kosugi, T., et~al. 2007, \solphys, 243, 3

\bibitem[{{Lin}, {Kuhn} \& {Coulter}(2004)}]{Lin2004}
Lin, H., Kuhn, J.~R. \& Coulter, R. 2004, \apj, 613, L177

\bibitem[{{Marsh} \& {Walsh}(2006)}]{Marsh}
Marsh, M.~S. \& Walsh, R.~W. 2006, \apj, 643, 540

\bibitem[{{McIntosh}, {De Pontieu} \& {Tomczyk}(2008)}]{McIntosh2008}
McIntosh, S.W., De Pontieu, B., Tomczyk, S. 2008, \solphys, 252, 321

\bibitem[{{Melrose}(1977)}]{Melrose}
Melrose, D.~B. 1977, Aust.J.Phys., 30, 495

\bibitem[{{Nakariakov} \& {Verwichte}(2005)}]{Nak2005}
Nakariakov, V.~M. \& Verwichte, E. 2005, Living Rev. Solar Phys., 2, 3

\bibitem[{{Nakariakov} {et~al.}(1999)}]{Nak1999}
Nakariakov, V.~M. Ofman, L., DeLuca, E.~E., Roberts, B \& Davila, J.~M. 1999, Science, 285, 862

\bibitem[{{Penn}(1994)}]{penn}
Penn, M.~J. 1994, Space Sci. Rev., 70, 185

\bibitem[{{Sakurai} {et~al.}(2002)}]{Sakurai}
Sakurai, T., Ichimoto, K., Raju, K.~P. \& Singh, J. 2002, \solphys, 209, 265

\bibitem[{{Schrijver}, {Aschwanden}, {Title}(2002)}]{Sch2002}
Schrijver, C.~J., Aschwanden, M.~J., \& Title, A.~M. 2002, \solphys, 206, 69

\bibitem[{{Singh}, {et~al.}(2002)}]{singh}
Singh, J., Sakurai, T., Ichimoto, K. \& Takeda, A. 1981, PASJ, 54, 807

\bibitem[{{Smartt}, {Dunn} \& {Fisher}(1981)}]{Smartt1981}
Smartt, R.N., Dunn, R.B., \& Fisher, R.R. 1981, Proc. SPIE, 288, 395

\bibitem[{{Suzuki} \& {Inutsuka}(2005)}]{Suzuki2005}
Suzuki, T. \& Inutsuka, S. 2005, \apj, 632, L49

\bibitem[{{Thompson} {et~al.}(1998)}]{Thompson}
Thompson, B.~J., Plunkett, S.~P., Gurman, J.~B., Newmark, J.~S., St Cyr, O.~C. \& Michels, D.~J. 1998, 
GRL, 25, 2465

\bibitem[{{Tomczyk} {et~al.}(1995)}]{Tomczyk1995}
Tomczyk, S., Streander, K., Card, G., Elmore, D., Hull, H. \& Cacciani, A. 1995, \solphys, 159, 1

\bibitem[{{Tomczyk} {et~al.}(2007)}]{Tomczyk2007}
Tomczyk, S., et~al. 2007, Science, 317, 1192

\bibitem[{{Tomczyk} {et~al.}(2008)}]{Tomczyk2008}
Tomczyk, S., et al. 2008, \solphys{}, 247, 411.

\bibitem[{{Van Doorsselaere}, {Nakariakov} \& {Verwichte}(2008)}]{Van2008a}
Van Doorsselaere, T., Nakariakov, V.~M., Verwichte, E. 2008, \apj, 676, L73

\bibitem[{{Van Doorsselaere} {et~al.}(2008)}]{Van2008b}
Van Doorsselaere, T., Nakariakov, V.~M., Young P.~R., Verwichte, E. 2008, A \& A, 487, L17

\bibitem[{{Van~Vleck}(1925)}]{vanvleck}
Van~Vleck, J,~H. 1925, Proc. Nat. Acad. Sci., 11, 612

\bibitem[{{Verdini} \& {Velli}(2007)}]{Verdini2007}
Verdini, A. \& Velli, M. 2007, \apj, 662, 701

\bibitem[{{Wilhelm} {et~al.}(1995)}]{Wilhelm1995}
Wilhelm, K., et~al. 1995, \solphys, 162, 189


\bibitem[{{Withbroe} \& {Noyes}(1977)}]{with}
Withbroe, G.~L. \& Noyes, R.~W. 1977, Ann. Rev. Astron. Astrophys, 15, 363

\bibitem[{{Young} {et~al.}(2008)}]{Young2008}
Young, P.~R., Watanabe, T., Hara, H. \& Mariska, J.~T., 2008, A \& A, accepted

\end{thebibliography}
\end{document}